\documentclass[pdflatex,iicol,sn-mathphys-num]{sn-jnl}
\usepackage{amsmath,amssymb,mathtools,bm,graphicx,booktabs,array}
\hypersetup{pdftitle={Mass-Weighted Coherence Zeros and Ordering-Dependent Survival Bounds in Three-Flavor Neutrinos},pdfauthor={Zhongze Guo and Bei Xu},pdfdisplaydoctitle=true}
\newcommand{\Tr}{\operatorname{Tr}}
\newcommand{\Ree}{\operatorname{Re}}
\newcommand{\Imm}{\operatorname{Im}}
\newcommand{\Id}{\mathbb I}

\begin{document}
\title{Mass-Weighted Coherence Zeros and Ordering-Dependent Survival Bounds in Three-Flavor Neutrinos}
\author[1]{\fnm{Zhongze} \sur{Guo}}\email{guozhongze007@gmail.com}
\author*[2]{\fnm{Bei} \sur{Xu}}\email{xubei0903@163.com}

\affil[1]{\orgdiv{Department of Physics and Institute of Theoretical Physics}, \orgname{University of Science and Technology Beijing}, \orgaddress{\city{Beijing}, \postcode{100083}, \country{China}}}
\affil[2]{\orgdiv{Institute for Advanced Study}, \orgname{Tsinghua University}, \orgaddress{\city{Beijing}, \postcode{100084}, \country{China}}}

\abstract{In standard ultrarelativistic neutrino propagation,
the squared-mass term in the Hamiltonian gives rise to a mass-weighted flavor amplitude. We investigate whether this amplitude can vanish through coherent interference while the
electron-neutrino survival probability remains large. We characterize this cancellation through a complex mass-weighted response, whose real part also appears in the strongly suppressed Higgs--weak interference kernel at leading nonzero mass order in a minimal Dirac mass extension of the Standard Model. We derive a triangle criterion for cancellation at fixed source populations and a sharp, propagation-independent probability bound valid for arbitrary incident density matrices.
At the adopted central oscillation parameters, inverted mass ordering limits the electron-flavor probability to about 2.3\% whenever the real part of the response vanishes,
irrespective of the lightest neutrino mass. In contrast, standard propagation through a model of layered continental crust realizes a complex response zero at a survival probability near 89\% for a normal-ordering benchmark, nearly saturating its theoretical bound. This pronounced ordering dependence originates from the relative dispersion of the squared masses in the electron-flavor state. Since ordinary flavor probabilities are insensitive to a
common mass-squared shift, testing this distinction requires additional response information. For a specified mass-dependent
neutrino--electron interaction, we establish a conditional two-bin recoil readout with calibrated couplings and rate normalization and independently known flavor probability.
We identify an exact degeneracy between the common mass-squared shift and flavor-universal couplings in the general interaction model.}
\maketitle

\section{Introduction}\label{sec:intro}
In standard three-flavor mixing~\cite{Pontecorvo1968,MNS1962}, the squared neutrino masses govern the leading mass-dependent corrections to the relativistic dispersion relation and enter the vacuum oscillation Hamiltonian. Atmospheric and solar neutrino observations established flavor conversion~\cite{SuperK1998,SNO2002}, and global fits constrain the mixing weights and mass-squared splittings~\cite{NuFIT60,NuFIT61}. While standard flavor probabilities are invariant under a common shift of the squared masses, direct kinematic measurements such as KATRIN probe the absolute mass scale~\cite{KATRIN2025}. Flavor projection of the mass-squared operator gives rise to a coherent mass-weighted amplitude whose cancellation need not coincide with a zero of the ordinary flavor amplitude. Here we investigate how the neutrino mass spectrum and flavor-mixing structure constrain such cancellation, particularly when the electron-neutrino survival probability remains large. Testing this distinction requires information beyond standard flavor probabilities.

We compare the ordinary and mass-weighted amplitudes
for the same incident state and the same projection
onto electron flavor. In the neutrino mass basis, let
$D=\operatorname{diag}(d_1,d_2,d_3)$ with
$d_i=m_i^2\ge0$, and let $u$ denote the electron-flavor
state with components $u_i=U_{ei}^*$ and projector
$P_e=uu^\dagger$. For a normalized density matrix
$\rho$ at detection, we define the mass-weighted
response and the electron-flavor probability by
\begin{equation}\label{eq:Q}
	Q_2[\rho]=\Tr(\rho DP_e),\qquad
	P_{ee}[\rho]=\Tr(\rho P_e).
\end{equation}
Here $P_{ee}$ is a survival probability for an initially
electron-flavor neutrino. The response is generally
complex because $D$ and $P_e$ do not commute.
Its real and imaginary parts are the expectations
of $\{D,P_e\}/2$ and $[D,P_e]/(2i)$, respectively.
For a pure state $\rho=|a\rangle\langle a|$, the ordinary
and mass-weighted amplitudes are
$A_e=\langle u|a\rangle$ and $B_2=\langle u|D|a\rangle$.
The postselected mass-squared weak value
$D_w=B_2/A_e$~\cite{AAV1988} then gives
\begin{equation}\label{eq:weakvalue}
	Q_2=B_2^*A_e=P_{ee}D_w^*,\qquad (P_{ee}>0).
\end{equation}
A non-nodal zero occurs when the mass-weighted
amplitude vanishes while the ordinary flavor amplitude
remains nonzero. Related mass-weighted amplitudes
enter the ultrarelativistic expansion of
flavor-postselected neutrino energy and momentum,
with their physical interpretation depending on
the production and detection
protocol~\cite{PortoSilva2021}.

A complementary physical origin of the squared-mass
weighting arises from neutrino interactions.
In a minimal Dirac mass extension of the Standard
Model, the Higgs coupling to a neutrino is
proportional to its mass. A further mass factor
arises from the chiral structure of the interference
between Higgs exchange and the standard weak amplitude.
For coherent, negative-helicity incident neutrinos
at common ultrarelativistic reference kinematics,
the charged-current part of this interference
contains a term proportional to $\Ree Q_2$ at
leading nonzero order (Sect.~\ref{sec:recoil}).
Although the Higgs contribution is too strongly
suppressed to provide a practical measurement,
it illustrates how the squared-mass-weighted
flavor response arises from a mass-dependent
interaction.

General weak-value theory provides constraints
on the postselection probability at a prescribed
weak value~\cite{PangDresselBrun2014}.
Here we investigate how the neutrino mass splittings,
absolute mass scale, and electron-flavor mixing
weights constrain the coherent cancellation of
the mass-squared amplitude.
We determine the full admissible response region
for arbitrary incident density matrices, derive
ordering-dependent restrictions on the electron-flavor
probability, and investigate their dynamical
realization under standard matter evolution.

The cancellation conditions depend on the coherence
of the neutrino state, which is governed by its
production, propagation, and detection
processes~\cite{GiuntiKimLee1991,GiuntiKimLee1992,
	GiuntiKimLee1993,AkhmedovKopp2010,
	AkhmedovKopp2010Erratum,AkhmedovWilhelm2013}.
For a coherent electron-neutrino source with fixed
mass-basis populations, cancellation requires the
three weighted contributions to form a closed
triangle in the complex plane.
Vacuum propagation restricts the relative phases
to a trajectory determined by $L/E$.
In matter, coherent forward scattering through
the Mikheyev--Smirnov--Wolfenstein (MSW)
effect~\cite{Wolfenstein1978,MikheyevSmirnov1985,
	Parke1986,KuoPantaleone1989} can change both the
mass-basis populations and coherences.
The mass-weighted amplitude is directly related
to the local derivative of the flavor amplitude
after removing the explicit matter contribution.
The cancellation conditions must therefore be
evaluated for the propagated state.

For terrestrial propagation, variations of the
baseline and energy can generate a two-parameter
family of propagated states in which isolated
response zeros may occur.
Changing the baseline generally selects a
different Earth chord and density
profile~\cite{BlennowSmirnov2013}.
We distinguish derivatives along a fixed
trajectory from those across different trajectories
and characterize the local winding of the
response near isolated zeros, rather than
interpreting it as an independent geometric
phase of the neutrino state.

We first establish the mass-dependent conditions
for coherent cancellation at fixed source populations
and quantify the coherence required for such
cancellation (Sect.~\ref{sec:source}).
We then examine the response under matter evolution
(Sects.~\ref{sec:matter} and \ref{sec:numerical}).
For a normal-ordering benchmark, standard three-flavor
propagation through a model of layered continental
crust realizes a complex response zero at a survival
probability near $89\%$, nearly saturating its
theoretical upper bound.
In contrast, for inverted ordering at the adopted
central oscillation parameters, the electron-flavor
probability cannot exceed about $2.3\%$ whenever
the real part of the response vanishes, irrespective
of the lightest neutrino mass or the propagation
history. This restriction holds for arbitrary
incident density matrices.
The pronounced ordering dependence originates
from the interplay between the mass-squared
spectrum and electron-flavor mixing weights.
Testing this distinction requires independent
information about the real part of the response.
Section~\ref{sec:recoil} establishes a conditional
recoil-based readout for a specified mass-dependent
neutrino--electron interaction and identifies the associated coupling degeneracy.

\section{Mass-weighted coherent selection rules}\label{sec:source}
\subsection{Cancellation at fixed source populations}
For a coherent electron-flavor source, the vacuum mass-basis populations are $w_i=|u_i|^2$, with $\sum_i w_i=1$. Fix the overall phase by setting $\theta_1=0$. Allowing the two relative phases to vary independently defines a torus of states with amplitudes $a_i(\bm\theta)=u_i e^{-i\theta_i}$. The survival and mass-weighted amplitudes on this torus are
\begin{equation}\label{eq:AB}
\begin{aligned}
A(\bm\theta)&=\sum_{i=1}^3w_i e^{-i\theta_i},\\
B_2(\bm\theta)&=\sum_{i=1}^3b_i e^{-i\theta_i},\\
b_i&=d_iw_i,\qquad Q_2=B_2^*A.
\end{aligned}
\end{equation}
The measured electron-row weights satisfy $w_1>1/2$, so the reverse triangle inequality gives $|A|\ge2w_1-1>0$. Every zero of $Q_2$ on the torus is therefore non-nodal.

Since $A\ne0$, the response vanishes precisely when the three complex contributions to $B_2$ sum to zero. For positive $b_i=d_iw_i$, cancellation is possible if and only if the lengths satisfy the triangle inequalities, $b_i\le b_j+b_k$ for each choice of distinct $i,j,k$. With the first contribution on the positive real axis, the cancellation condition becomes
\begin{equation}\label{eq:triangle_phases}
\begin{aligned}
b_1+b_2\cos\theta_2+b_3\cos\theta_3&=0,\\
b_2\sin\theta_2+b_3\sin\theta_3&=0.
\end{aligned}
\end{equation}
The two relative phases then satisfy
\begin{equation}\label{eq:triangle_cos}
\cos\theta_2=\frac{b_3^2-b_1^2-b_2^2}{2b_1b_2},\qquad
\cos\theta_3=\frac{b_2^2-b_1^2-b_3^2}{2b_1b_3}.
\end{equation}
When the inequalities are strict, the triangle has exactly two configurations modulo $2\pi$, related by reflection about the real axis. The Jacobian determinant of the map from the two phases to $B_2$ is
\begin{equation}\label{eq:triangle_jac}
J_B=\det\frac{\partial(\Ree B_2,\Imm B_2)}{\partial(\theta_2,\theta_3)}
=b_2b_3\sin(\theta_2-\theta_3).
\end{equation}
At $B_2=0$, multiplication by $A$ preserves orientation and complex conjugation reverses it, so $J_Q=-|A|^2J_B$. Reflection changes the sign of $J_B$. The two zeros are therefore simple and have opposite indices. When the longest side equals the sum of the other two, the two configurations merge into a single collinear solution with $J_B=0$. If one side exceeds the sum of the other two, cancellation is impossible.

For normal ordering, $d_2=d_1+\Delta_{21}$ and $d_3=d_1+\Delta_{31}$, where $\Delta_{i1}=m_i^2-m_1^2$. At the benchmark mixing parameters, $\Delta_{31}w_3>\Delta_{21}w_2$, so $b_2<b_1+b_3$ for all $d_1>0$. The remaining inequalities $b_3<b_1+b_2$ and $b_1<b_2+b_3$ give the lower and upper bounds on $d_1$, respectively,
\begin{equation}\label{eq:mass-window}
\begin{aligned}
d_{1,\rm low}&=\frac{\Delta_{31}w_3-\Delta_{21}w_2}{1-2w_3},\\
d_{1,\rm high}&=\frac{\Delta_{21}w_2+\Delta_{31}w_3}{2w_1-1}.
\end{aligned}
\end{equation}
For the NuFIT 6.1 central parameters in Sect.~\ref{sec:numerical}, the interval for two simple zeros is approximately $5.94<m_1/{\rm meV}<15.01$.

For fixed electron-row weights and mass splittings, the lightest mass sets all three side lengths. Outside the closed interval no choice of the two relative phases gives closure. Within the open interval the two reflected phase configurations give $B_2=0$ at finite survival probability. Each local index specifies how $Q_2$ winds around its zero. By Eq.~\eqref{eq:weakvalue}, the postselected mass-squared weak value also vanishes there. In vacuum both relative phases depend on $L/E$, so a physical trajectory explores only a curve on the torus. In matter the mass-basis populations can change, and the interval in Eq.~\eqref{eq:mass-window} no longer bounds the masses at which a trajectory can reach a zero. Inferring the absolute mass would require an interaction sensitive to the response, as discussed in Sect.~\ref{sec:geometry}.

\subsection{Coherence conditions at detection}
For a density matrix $\rho$ at the detector with $P_{ee}>0$, the mixed-state postselected weak value is $\Tr(P_eD\rho)/P_{ee}=Q_2^*/P_{ee}$~\cite{DresselJordan2012}. A bound on the response itself applies at any $P_{ee}$. To obtain it, define
\begin{equation}\label{eq:coherence}
\begin{aligned}
q_{\rm diag}&=\sum_i d_iw_i\rho_{ii},\\
C_\rho&=\sum_{i<j}|\rho_{ij}|\sqrt{w_iw_j}(d_i+d_j).
\end{aligned}
\end{equation}
Writing $z_{ij}=\rho_{ij}u_j u_i^*$ for $i<j$ and grouping each pair of off-diagonal terms gives
\begin{equation}\label{eq:Qpair}
Q_2=q_{\rm diag}+\sum_{i<j}\big[(d_i+d_j)\Ree z_{ij}
+i(d_j-d_i)\Imm z_{ij}\big].
\end{equation}
Since $\Ree z_{ij}\ge-|z_{ij}|$ and $|z_{ij}|=|\rho_{ij}|\sqrt{w_iw_j}$,
\begin{equation}\label{eq:exclusion}
\Ree Q_2\ge q_{\rm diag}-C_\rho.
\end{equation}
The populations supply the nonnegative contribution $q_{\rm diag}$, and $C_\rho$ bounds the magnitude of the interference contribution to the real part. If $C_\rho<q_{\rm diag}$, the real part is strictly positive and a response zero is excluded. At equality, a zero requires saturation of the real-part bound and cancellation of the imaginary part. The criterion uses the state at the detector, since MSW propagation can generate off-diagonal terms from a mass-diagonal source.

To quantify the effect of partial coherence at fixed populations, apply a mass-basis dephasing channel to the torus states. The channel retains a fraction $\gamma$ of each off-diagonal element,
$\rho_\gamma=(1-\gamma)\operatorname{diag}(w_i)+\gamma|a(\bm\theta)\rangle\langle a(\bm\theta)|$, $0\le\gamma\le1$.
This convex mixture of complete dephasing and the identity channel is completely positive and gives
\begin{equation}\label{eq:torus_dephase_real}
\begin{aligned}
\Ree Q_2[\rho_\gamma]={}&h_0+\gamma\big[c_{12}\cos\theta_2+c_{13}\cos\theta_3\\
&\hphantom{h_0+\gamma\big[}{}+c_{23}\cos(\theta_2-\theta_3)\big],
\end{aligned}
\end{equation}
where $h_0=\sum_i d_iw_i^2$ and $c_{ij}=(d_i+d_j)w_iw_j$. For the benchmark masses and mixing weights,
\begin{equation}\label{eq:torus_condition}
c_{23}(c_{12}+c_{13})<c_{12}c_{13}.
\end{equation}
Let $F(\theta_2,\theta_3)$ denote the bracket in Eq.~\eqref{eq:torus_dephase_real} and set $x_i=|\cos(\theta_i/2)|$ for $i=2,3$. The half-angle identities and $|\sin[(\theta_2-\theta_3)/2]|\le x_2+x_3$ give
\begin{equation}\label{eq:halfangle_bound}
\begin{aligned}
\tfrac12\big[F(\theta_2,\theta_3)-F(\pi,\pi)\big]
&\ge c_{12}x_2^2+c_{13}x_3^2\\
&\quad-c_{23}(x_2+x_3)^2\ge0,
\end{aligned}
\end{equation}
where the last inequality follows from weighted Cauchy--Schwarz and Eq.~\eqref{eq:torus_condition}. Thus $F$ attains its global minimum at $(\pi,\pi)$, where the imaginary part of $Q_2$ also vanishes, and the first zero as $\gamma$ increases from zero occurs at
\begin{equation}\label{eq:gamma_critical}
\begin{aligned}
\gamma_c&=\frac{h_0}{c_{12}+c_{13}-c_{23}}\\
&=0.6164\quad\text{(benchmark; rounded).}
\end{aligned}
\end{equation}
For $\gamma<\gamma_c$, $\Ree Q_2>0$ at every point on the torus. This threshold describes dephasing of the fixed-population source states under condition~\eqref{eq:torus_condition}. The effect of dephasing during matter propagation is treated separately in Sect.~\ref{sec:numerical}.

For an arbitrary analysis operator $O$, the numerical range of $DO$ determines whether any density matrix can give a zero. Writing $\Ree K=(K+K^\dagger)/2$ for the Hermitian part of a matrix, we have
\begin{equation}\label{eq:numericalrange}
\begin{gathered}
\Tr(\rho DO)\ne0\quad\text{for all density matrices }\rho\\
\Longleftrightarrow\\
\exists\alpha\in\mathbb R:\ \Ree(e^{-i\alpha}DO)\succ0.
\end{gathered}
\end{equation}
The expectations of $K=DO$ form its compact convex numerical range. If this set excludes the origin, a separating line places it in an open half-plane, which gives the positive-definiteness condition in Eq.~\eqref{eq:numericalrange}. Conversely, a positive-definite Hermitian part has positive expectation in every state. For $O=P_e$, a state orthogonal to electron flavor gives a trivial zero at $P_{ee}=0$. The bound in Sect.~\ref{sec:numerical} addresses zeros at fixed, nonzero electron-flavor probability.

\section{MSW evolution, profile geometry, and the derivative identity}\label{sec:matter}
\subsection{Exact identity along one fixed propagation trajectory}
Along a fixed trajectory with piecewise continuous matter potential $V(\ell)$, the coherent amplitude $a$ in the vacuum mass basis obeys the MSW evolution equation~\cite{Wolfenstein1978,MikheyevSmirnov1985,Parke1986,KuoPantaleone1989,BlennowSmirnov2013}
\begin{equation}\label{eq:schrod}
i\frac{d a}{d\ell}=\left[\frac{D}{2E}+V(\ell)P_e\right]a,\qquad a(0)=u,
\end{equation}
where $\ell$ is the distance along the trajectory in $\mathrm{eV}^{-1}$ and $u$ is the electron-flavor source state. Projecting onto $u$ gives the evolution of the survival amplitude $A_e(\ell)=u^\dagger a(\ell)$ in terms of $B_2(\ell)=u^\dagger D a(\ell)$,
\begin{equation}\label{eq:projected_schrod}
i\partial_\ell A_e=\frac{B_2}{2E}+V(\ell)A_e,
\qquad B_2=2iE(\partial_\ell+iV)A_e.
\end{equation}
For this pure state, $Q_2=\Tr(|a\rangle\langle a|DP_e)$. Substituting for $B_2$ gives
\begin{equation}\label{eq:local_identity}
\begin{aligned}
Q_2&=B_2^*A_e\\
&=-2iEA_e\left[(\partial_\ell+iV(\ell))A_e\right]^*.
\end{aligned}
\end{equation}
Where $A_e\ne0$, write $A_e=\sqrt{P_{ee}}e^{i\varphi_e}$. The real and imaginary parts of Eq.~\eqref{eq:local_identity} are then
\begin{equation}\label{eq:phase}
\begin{aligned}
\Ree Q_2&=-2EP_{ee}(\partial_\ell\varphi_e+V),\\
\Imm Q_2&=-E\,\partial_\ell P_{ee}.
\end{aligned}
\end{equation}
At a non-nodal zero, the matter-covariant derivative $(\partial_\ell+iV)A_e$ vanishes. Equivalently, $P_{ee}$ is stationary along the trajectory and $\partial_\ell\varphi_e=-V$ in the phase convention of Eq.~\eqref{eq:schrod}. Together these conditions constrain the local evolution even when $P_{ee}$ is large. Equation~\eqref{eq:projected_schrod} thus identifies $B_2/(2E)$ as the mass contribution to the local evolution of the flavor amplitude. If the propagation Hamiltonian is written with $D-d_{\rm ref}\Id$ for a constant $d_{\rm ref}$, the response inferred from the derivative in that convention is $Q_2-d_{\rm ref}P_{ee}$. Recovering the response defined in Eq.~\eqref{eq:Q} requires restoring this offset. Standard flavor probabilities are insensitive to it. At a density jump the corresponding derivatives are one-sided. For an isolated non-nodal zero in a two-parameter family, the complex conjugation in Eq.~\eqref{eq:local_identity} reverses the winding of the derivative. This index characterizes the response around the zero (Sect.~\ref{sec:geometry}).

\subsection{Baseline derivatives across Earth chords}\label{sec:chordderiv}
Earth oscillograms describe amplitude and phase conditions across the baseline--energy plane~\cite{AkhmedovMaltoniSmirnov2007}. Changing the length $L$ of an Earth chord also changes the depths sampled along the path. Let $s$ denote the distance from the source in km and $V(s;L)$ the matter potential along that chord. The endpoint survival amplitude is $A_e(L;L)=u^\dagger U_L(L,0)u$, where $U_L$ is the propagator for the profile labeled by $L$. With $H(s;L)=D/(2E)+V(s;L)P_e$ and $\kappa=5.067730716\times10^9\,{\rm eV}^{-1}{\rm km}^{-1}$, differentiating at fixed energy gives
\begin{equation}\label{eq:chordderivative}
\begin{aligned}
\frac{d}{dL}A_e(L;L)={}&-i\kappa\,u^\dagger H(L;L)U_L(L,0)u \\
&-i\kappa\int_0^L\!ds\,u^\dagger U_L(L,s)\\
&\quad\times \frac{\partial H(s;L)}{\partial L}U_L(s,0)u.
\end{aligned}
\end{equation}
The first term is the derivative with respect to the upper endpoint at fixed profile. For the second term, introduce an independent profile label $\lambda$. Varying this label with the endpoints fixed at $0$ and $L$ gives the Duhamel identity
\begin{equation}\label{eq:duhamel}
\begin{aligned}
&\left.\partial_\lambda U_\lambda(L,0)\right|_{\lambda=L}\\
&\quad=-i\kappa\int_0^L ds\,U_L(L,s)
(\partial_L H)U_L(s,0).
\end{aligned}
\end{equation}
For a sharp spherical interface, $\partial_L H$ contains a delta function at each moving crossing. Differentiating the ordered product of the layer propagators gives the same contribution (see \hyperref[app:earth]{Appendix~\ref*{app:earth}}). At a coherent non-nodal zero, Eq.~\eqref{eq:phase} sets the endpoint contribution to $dP_{ee}/dL$ to zero. The change in profile then gives the entire slope across chords,
\begin{multline}\label{eq:non-extremum}
\left.\frac{d}{dL}P_{ee}(L;L)\right|_{Q_2=0}
\\ =2\Ree\biggl\{A_e^*\biggl[-i\kappa\int_0^L\!ds\,u^\dagger U_L(L,s)\\
\times \partial_L H(s;L)U_L(s,0)u\biggr]\biggr\},
\end{multline}
which can be nonzero even though $P_{ee}$ is stationary along the fixed trajectory. In a constant-density slab whose density is independent of $L$, the integrand vanishes.

Denote the contribution to $dA_e/dL$ from varying the potential by
\begin{equation}\label{eq:path_term}
\begin{aligned}
\mathcal W_e(L,E)&=-i\kappa\int_0^L\!ds\,u^\dagger U_L(L,s)\\
&\quad\times [\partial_LV(s;L)]P_eU_L(s,0)u.
\end{aligned}
\end{equation}
Here $\partial_L V$ is taken at fixed $s$ and $E$. Using $dP_{ee}/dL=2\Ree(A_e^*dA_e/dL)$ gives
\begin{equation}\label{eq:chord_slope}
\frac{dP_{ee}}{dL}\bigg|_{\rm chords}
=-\frac{\kappa}{E}\Imm Q_2
+2\Ree[A_e^*\mathcal W_e].
\end{equation}
The endpoint term contains $\Imm Q_2$, with $E$ expressed in eV. The second term accumulates the change in potential along the chord. For a smooth profile family, unitarity and $\|P_e\|=1$ bound the magnitude of this term by
\begin{equation}\label{eq:chord_bound}
\big|2\Ree(A_e^*\mathcal W_e)\big|
\le 2\kappa\int_0^L\!ds\,|\partial_LV(s;L)|.
\end{equation}
For sharp shells, the right-hand side uses the total variation of the signed measure $\partial_L V$. The ordered-layer derivative is well defined near each simple zero considered here. At a chord tangent to a shell boundary, the crossing position can have a singular derivative with respect to $L$, so Eq.~\eqref{eq:chord_bound} gives no uniform smallness estimate there. Both contributions to the slope in Eq.~\eqref{eq:chord_slope} follow from the same MSW Hamiltonian.

\section{Response zeros in matter and the ordering-dependent bound}\label{sec:numerical}
\subsection{Zeros in model crustal profiles}
We propagate an electron neutrino through matter to locate zeros of $Q_2$ in the baseline--energy plane. We examine how the root position and local index respond to changes in the matter profile, oscillation parameters, and coherence, and evaluate the across-chord probability slope at the root. For coherent propagation we use the standard unitary three-flavor MSW Hamiltonian~\cite{BlennowSmirnov2013}, with $R_\oplus=6371$ km, $Y_e=0.5$, and $V=7.56\times10^{-14}\,\mathrm{eV}\,(\rho_m/\mathrm{g\,cm^{-3}})Y_e$. We consider uniform matter, surface shells based on the Preliminary Reference Earth Model (PREM)~\cite{PREM1981}, a layered continental crust, and a crust with a piecewise linear density profile. \hyperref[app:earth]{Appendix~\ref*{app:earth}} specifies these illustrative profiles.

The propagation benchmark uses normal ordering with $m_1=7$ meV. We take the central values from the NuFIT 6.1 (November 2025) IC24 fit with Super-Kamiokande atmospheric data. These are $\sin^2\theta_{12}=0.3088$, $\sin^2\theta_{13}=0.02248$, $\Delta m^2_{21}=7.537\times10^{-5}\,\mathrm{eV}^2$, and $\Delta m^2_{31}=2.511\times10^{-3}\,\mathrm{eV}^2$~\cite{NuFIT61,NuFIT60}. The NuFIT quantity $\Delta m^2_{3\ell}$ equals $\Delta m^2_{31}$ for this ordering. In standard electron matter, $P_{ee}$ is independent of $\theta_{23}$ and $\delta_{\rm CP}$. We compute a matrix exponential for each constant-density layer and use a midpoint discretization for the depth-gradient profile. At each benchmark root, we evaluate the Jacobian determinant of $(\Ree Q_2,\Imm Q_2)$ with respect to $(L,E)$ and independently compute the winding on a counterclockwise loop around the root. OpenAI Codex (GPT-6) assisted with the derivation of the ordering bound, the spinor calculation in Sect.~\ref{sec:recoil}, and numerical checks of the propagation and recoil results. The numerical files available from the corresponding author document the checks against density matrices that attain the bound and provide the propagation inputs, root coordinates, and convergence tests.

\begin{table}[tbp]
\caption{Baseline and energy at simple non-nodal zeros of $Q_2$ for the normal-ordering NuFIT 6.1 benchmark ($m_1=7$ meV, $\Gamma=0$). Each zero has local index $-1$.}
\label{tab:roots}
\centering\begin{tabular}{@{}lrr@{}}
\toprule
Matter profile & $L_0$ (km) & $E_0$ (GeV)\\
\midrule
Uniform $2.8\,\mathrm{g/cm^3}$ & 678.4 & 1.2173\\
Layered continental & 687.5 & 1.2354\\
Depth-gradient crust & 681.5 & 1.2244\\
\bottomrule
\end{tabular}
\end{table}

\begin{figure*}[t]
\centering\includegraphics[width=\textwidth]{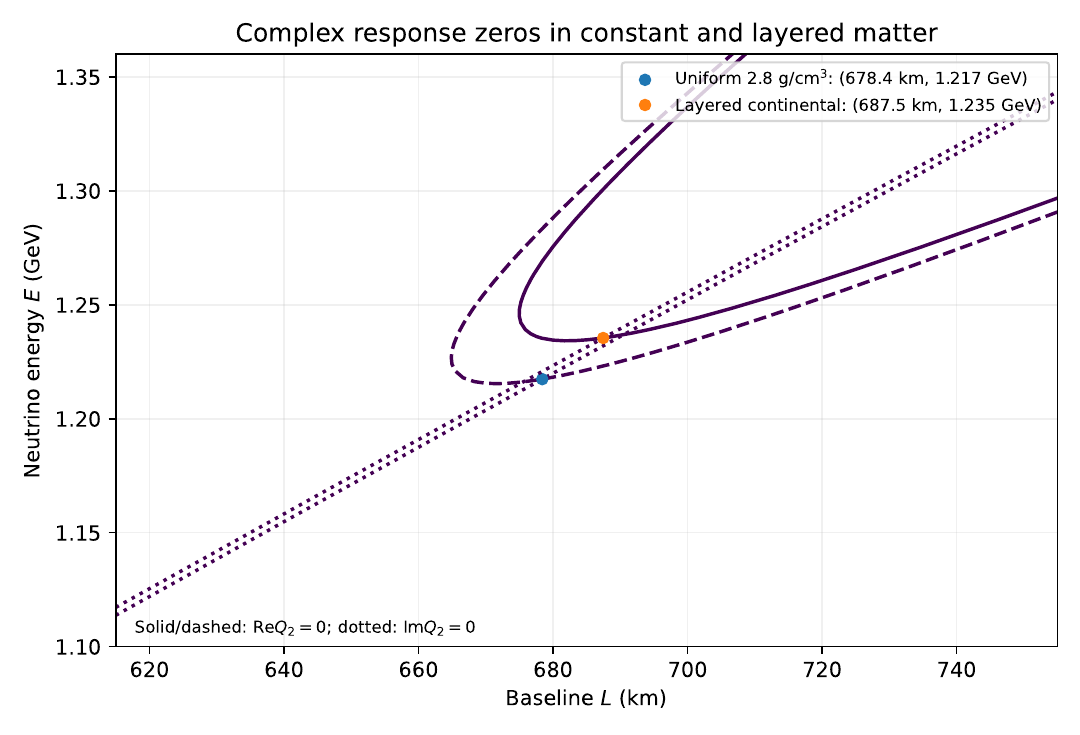}
\caption{Zero contours of the complex response for the normal-ordering benchmark with $m_1=7$ meV and the NuFIT 6.1 central oscillation parameters. Solid and dashed curves mark $\Ree Q_2=0$ for layered continental and uniform matter, respectively. Dotted curves mark $\Imm Q_2=0$. Each marker identifies the intersection for one profile, where $Q_2=0$ and $P_{ee}>0$}\label{fig:contours}
\end{figure*}

All three profiles in Table~\ref{tab:roots} yield a simple zero. At the continental root in Fig.~\ref{fig:contours}, the electron-neutrino survival probability is $P_{ee}\simeq0.88969$, the Jacobian determinant is negative, and the numerical residual $|Q_2|$ is of order $10^{-18}\,\mathrm{eV}^2$. The corresponding trajectory crosses the density interface at a depth of $5$ km. The PREM-inspired profile gives $(L_0,E_0)\simeq(730.6\,\mathrm{km},1.3110\,\mathrm{GeV})$ on a chord that lies entirely in its constant-density surface shell. We also find a simple root of index $-1$ at each of 21 sampled interpolations between uniform matter at $2.8\,\mathrm{g/cm^3}$ and the layered continental profile.

\subsection{Upper bound on the electron-flavor probability}
For any incident density matrix $\rho$, write $p=\Tr(\rho P_e)$ and define the electron-row mean $\mu=\sum_i w_i d_i$ and variance $\sigma^2=\sum_i w_i(d_i-\mu)^2$ of the squared masses. The measured splittings and mixing weights give $\sigma>0$, so $v=(D-\mu\Id)u/\sigma$ is a unit vector orthogonal to $u$. It follows that $Q_2=u^\dagger\rho Du=\mu p+\sigma u^\dagger\rho v$. Positivity and unit trace give $|u^\dagger\rho v|^2\le p\,v^\dagger\rho v\le p(1-p)$, hence
\begin{equation}\label{eq:ordering_disk}
 |Q_2-\mu p|\le\sigma\sqrt{p(1-p)}.
\end{equation}
At fixed $p$, every point in this disk is attained by a density matrix supported on the span of $u$ and $v$, with diagonal entries $(p,1-p)$ and the corresponding off-diagonal entry. The center $\mu p$ is real, so the disk intersects the imaginary axis precisely when it contains the origin. Thus the largest $p$ compatible with $\Ree Q_2=0$ is also the largest $p$ compatible with $Q_2=0$,
\begin{equation}\label{eq:ordering_ceiling}
 p_*=\frac{\sigma^2}{\mu^2+\sigma^2}.
\end{equation}
For a pure state with $P_{ee}>0$, a zero requires $a$ to be orthogonal to $Du$. The maximizing state is the normalized projection of $u$ onto that orthogonal subspace. This gives the zero-weak-value case of the optimal postselection probability~\cite{PangDresselBrun2014}, with the roles of the fixed and varied states interchanged.

Equation~\eqref{eq:ordering_ceiling} is controlled by the relative dispersion $\sigma/\mu$ of the squared masses in electron flavor. In normal ordering, most of the electron-flavor weight resides in the lighter states, while the heaviest state can make an appreciable contribution to the mass-weighted amplitude despite its small mixing weight. At fixed splittings and mixing weights, increasing the lightest mass squared raises $\mu$ by the same amount and leaves $\sigma^2$ unchanged. In inverted ordering, the two nearly degenerate heavy states carry almost all of the electron-flavor weight. The ceiling is largest at $m_3=0$. There, the NuFIT 6.1 IC24 central values with Super-Kamiokande atmospheric data ($\sin^2\theta_{12}=0.3088$, $\sin^2\theta_{13}=0.02262$, $\Delta m^2_{21}=7.537\times10^{-5}\,\mathrm{eV}^2$, and $\Delta m^2_{32}=-2.483\times10^{-3}\,\mathrm{eV}^2$) give $p_*=0.022821$~\cite{NuFIT61}. For the normal-ordering benchmark at $m_1=7$ meV, $p_*=0.893014$. The continental trajectory reaches $P_{ee}=0.889693$ at its complex zero, or $99.63\%$ of the maximum over all incident states at these parameters. Equation~\eqref{eq:ordering_ceiling} holds independently of the propagation history. Its use with a measured survival probability requires independent evidence that $\Ree Q_2=0$.

\begin{figure*}[t]
\centering\includegraphics[width=\textwidth]{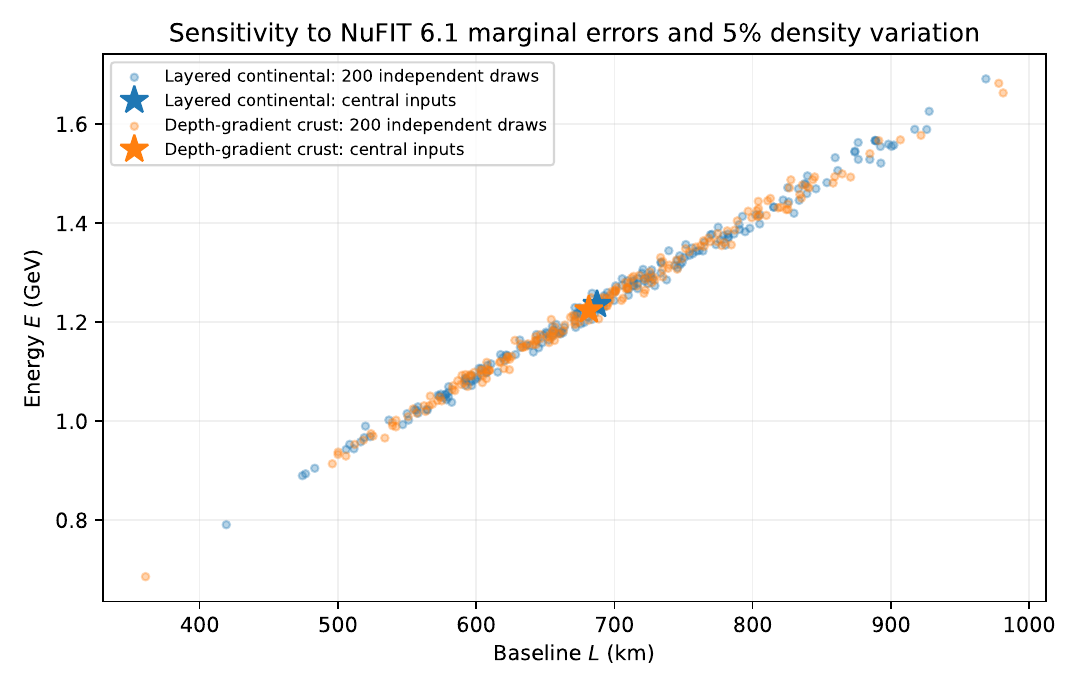}
\caption{Positions of the simple zero for the continental and depth-gradient profiles. Each cloud comprises 200 independent draws using the marginal oscillation-parameter errors and a 5\% density-normalization uncertainty. The draws omit correlations in the oscillation fit. Stars denote the roots at the central inputs}\label{fig:uncertainties}
\end{figure*}

\subsection{Parameter variations and partial coherence}
For Fig.~\ref{fig:uncertainties} we draw $\sin^2\theta_{12}$, $\sin^2\theta_{13}$, $\Delta m^2_{21}$, and $\Delta m^2_{31}$ independently. We approximate the asymmetric marginal errors of the NuFIT 6.1 normal-ordering fit~\cite{NuFIT61} by Gaussian standard deviations of $0.0067$, $0.00057$, $1.0\times10^{-6}\,\mathrm{eV}^2$, and $2.1\times10^{-5}\,\mathrm{eV}^2$, respectively. For each draw, one independent Gaussian factor with mean 1 and standard deviation 0.05 scales the entire density profile. All sampled factors exceed the lower cutoff of 0.6 used in the code. All 200 draws for each profile yield a simple root of index $-1$. For the continental profile the median baseline is $693$ km. The marginal $2.5$th--$97.5$th percentile ranges are $508$--$901$ km in baseline and $0.94$--$1.57$ GeV in energy. These ranges describe the independent draws and omit correlations in the global fit. Across all 400 draws, $P_{ee}/p_*$ at the root ranges from $99.56\%$ to $99.70\%$, with $p_*$ evaluated at each draw's parameters. Thus the survival probability remains close to its ceiling as the root position varies. For a smooth variation of a parameter $t$, the local displacement of the root $x_0=(L_0,E_0)$ obeys $dx_0/dt=-J^{-1}\partial_tF$, where $F=(\Ree Q_2,\Imm Q_2)^T$ and $J=\partial_xF$ is nonsingular at $x_0$.

To test sensitivity to partial coherence, we introduce a phenomenological dephasing channel in the vacuum mass basis~\cite{BenattiFloreanini2000,Lisi2000,Gago2001,DeRomeri2023},
\begin{equation}\label{eq:lindblad}
\begin{aligned}
\frac{d\rho}{ds}&=-i[\kappa H,\rho]-\Gamma(\rho-\Delta\rho),\\
\Delta\rho&=\sum_i|i\rangle\langle i|\rho|i\rangle\langle i|,
\end{aligned}
\end{equation}
where $s$ is the distance along a fixed trajectory in km and $\rho(0)=P_e$. For $\Gamma\ge0$ the jump operators $\sqrt{\Gamma}|i\rangle\langle i|$ generate completely positive, trace-preserving evolution and damp the off-diagonal mass-basis coherences. In the continental profile, $\Gamma=10^{-6}$ and $10^{-5}\,\mathrm{km}^{-1}$ shift the zero to $(688.6\,\mathrm{km},1.2373\,\mathrm{GeV})$ and $(699.5\,\mathrm{km},1.2547\,\mathrm{GeV})$, respectively. Both roots retain index $-1$. Comparison with experimental constraints requires matching the dephasing operators and their energy dependence. Along a fixed path, differentiating $P_{ee}=\Tr(\rho P_e)$ and using $\Tr(\rho[P_e,D])=-2i\,\Imm Q_2$ gives
\begin{equation}\label{eq:mixedslope}
\frac{dP_{ee}}{ds}=-\frac{\kappa}{E}\Imm Q_2-\Gamma\left(P_{ee}-\sum_i\rho_{ii}w_i\right),
\end{equation}
where $E$ is in eV and $\kappa$ converts km to $\mathrm{eV}^{-1}$.

At the coherent continental root, the endpoint contribution $-(\kappa/E)\Imm Q_2$ to the probability slope vanishes. Varying the baseline at fixed energy also changes the chord profile, giving the benchmark total slope
\begin{equation}\label{eq:measuredslope}
\left.\frac{dP_{ee}}{dL}\right|_{\rm chords,\ Q_2=0}
=-2.52\times10^{-6}\,\mathrm{km}^{-1}.
\end{equation}
At the coherent root in the depth-gradient profile the corresponding slope is $-5.03\times10^{-6}\,\mathrm{km}^{-1}$. In each case the nonzero slope comes from the profile-variation term in Eq.~\eqref{eq:chord_slope}, derived in Sect.~\ref{sec:chordderiv}.

\section{Geometric-phase comparison and physical information limits}\label{sec:geometry}
\subsection{Winding of the response near a zero}
The Pancharatnam phase is defined through overlaps between state vectors~\cite{Pancharatnam1956,Mehta2009}. Gauge-covariant constructions separate geometric and dynamical phases in neutrino evolution~\cite{Johns2022}. Nonlinear neutrino refraction can give rise to geometric phases~\cite{JohnsFuller2017}. For mixed states the phase also depends on the chosen prescription for transport~\cite{Uhlmann1986,Joshi2020}. The response $Q_2$ is evaluated at the propagation endpoint. A small loop in $(L,E)$ around an isolated zero samples a family of endpoints. The local index is the winding number of $Q_2$ about the origin as the loop is traversed counterclockwise. It measures a property of the response and does not require parallel transport around a closed path in state space.

At nonzero survival probability, Eq.~\eqref{eq:weakvalue} relates the response to flavor postselection~\cite{PortoSilva2021}. Equation~\eqref{eq:local_identity} expresses the same response in terms of the calculated survival amplitude in a fixed common-phase convention. The triangle criterion and the Earth-chord examples identify, respectively, relative phases that permit cancellation at fixed source populations and matter trajectories that realize it.

\subsection{Flavor probabilities under a common mass shift}
A common shift $D\to D+c\Id$ adds $c\Id/(2E)$ to the standard MSW Hamiltonian. This scalar term commutes with $H(s)$ everywhere along the path, so
\begin{equation}\label{eq:shift_propagator}
U_c(L,0)=e^{-ic\ell/(2E)}U(L,0),
\end{equation}
where $\ell$ is the total path length in natural units. The phase in Eq.~\eqref{eq:shift_propagator} cancels from the propagated density matrix. The same invariance holds in the dephasing model of Eq.~\eqref{eq:lindblad}, since $[c\Id,\rho]=0$ and the mass-basis jump projectors are fixed. The operator $DP_e$, however, acquires an additive term $cP_e$. Hence
\begin{equation}\label{eq:shift}
\rho^{(c)}=\rho,\qquad P_{ee}^{(c)}=P_{ee},\qquad Q_2^{(c)}=Q_2+cP_{ee}.
\end{equation}
Equation~\eqref{eq:shift} holds pointwise in $(L,E)$. Measurements of fixed flavor projectors over any set of baselines and energies therefore leave the common mass shift undetermined, even though it changes the real part of $Q_2$ wherever $P_{ee}>0$.

\subsection{Conditional extraction of the real part from electron recoils}\label{sec:recoil}
Consider elastic scattering on a free, unpolarized electron at rest. In the minimal Dirac extension of the Standard Model, the neutrino Yukawa interaction is $-h\sum_j(m_j/v)\bar\nu_j\nu_j$, with $v$ the electroweak vacuum expectation value~\cite{PDG2024EW}. General calculations of neutrino--electron scattering with scalar and vector couplings include finite-neutrino-mass terms~\cite{Marquez2024}. Here the scalar vertex is diagonal in mass. For an outgoing mass state $j$, the weak amplitude can receive contributions from any incident component $i$, while the Higgs amplitude receives only the incident component $j$.

Let $\xi_i$ denote an incident negative-helicity Dirac spinor and $\xi'_j$ an outgoing spinor with four-momentum $p'_j$. Write $\xi_{iL,R}=P_{L,R}\xi_i$. Summing over the outgoing spin gives
\begin{equation}\label{eq:scalar_spin}
\begin{aligned}
\mathcal S^\mu_{ji}
 &=\sum_{s'}(\bar\xi'_j\gamma^\mu P_L\xi_i)(\bar\xi'_j\xi_j)^*\\
 &=\bar\xi_j(\not{p}'_j+m_j)\gamma^\mu P_L\xi_i\\
 &=\bar\xi_{jR}\not{p}'_j\gamma^\mu\xi_{iL}
   +m_j\bar\xi_{jL}\gamma^\mu\xi_{iL}.
\end{aligned}
\end{equation}
In the chiral basis the incident spinor satisfies $\|\xi_{jR}\|/\|\xi_{jL}\|=m_j/(E_j+|\mathbf p_j|)$. Thus both terms in the last line carry one factor of $m_j$. Evaluating the remaining spinors and momenta at common ultrarelativistic reference kinematics leaves the same leading kinematic coefficient for all $i,j$. The forward limit of a diagonal channel confirms that this coefficient is nonzero, since $\mathcal S^0_{jj}=2m_j\bar\xi_{jL}\gamma^0\xi_{jL}$ when $p'_j=p_j$. The scalar Yukawa vertex supplies the second factor of $m_j$. For coherent incident amplitudes $a_i$, the charged-current flavor factor is therefore
\begin{equation}\label{eq:scalar_flavor}
\sum_{j,i} a_j^*m_j^2(P_e)_{ji}a_i
  =a^\dagger DP_ea=Q_2 .
\end{equation}
For an unpolarized electron at rest, the scalar--vector electron spin sum is proportional to $(k+k')_\mu$, where $k,k'$ are the electron four-momenta. The tree-level kinematic coefficient is real because the incident helicity axis and the momenta lie in the scattering plane. The charged-current interference in the rate therefore selects $\Ree Q_2$. The flavor-universal neutral current gives $a^\dagger Da$. Averaging over an incident ensemble gives $\Ree Q_2[\rho]$ and $\Tr(\rho D)$ for the two flavor factors. The Yukawa couplings and the Higgs propagator strongly suppress this contribution.

To examine what a recoil spectrum could determine, we now specify a separate effective vector interaction at detection. Gauge-invariant constructions of nonstandard neutrino interactions and the bounds on their coefficients are discussed in Refs.~\cite{Gavela2009,Biggio2009,Ohlsson2013,FarzanTortola2018}. We take the corrections to the electron couplings to be proportional to $D$ in the vacuum mass basis and derive the dependence of the recoil spectrum on $\Ree Q_2$. For elastic scattering of a relativistic neutrino on a free, unpolarized electron at rest, the tree-level differential cross section is~\cite{Coloma2023ES}
\begin{equation}\label{eq:positive}
\begin{aligned}
\frac{d\sigma}{dT}&=\frac{2G_F^2m_e}{\pi}\Tr(\rho R),\\
R&=G_L^2+fG_R^2-\frac h2\{G_L,G_R\},\\
&=\left(G_L-\frac h2G_R\right)^2
+\left(f-\frac{h^2}{4}\right)G_R^2\succeq0,\\
&f=(1-T/E)^2,\qquad h=m_eT/E^2,
\end{aligned}
\end{equation}
where $T$ is the recoil kinetic energy, and neutrino masses are neglected in the external kinematics. Over the physical recoil range $0\le T\le2E^2/(2E+m_e)$, one has $1-T/E\ge m_eT/(2E^2)\ge0$. Squaring gives $f-h^2/4\ge0$, so the sum of squares in Eq.~\eqref{eq:positive} is positive semidefinite for any Hermitian $G_L,G_R$, including noncommuting couplings. We take
\begin{equation}\label{eq:G}
G_L=g_L\Id+P_e+\alpha_L D,\qquad G_R=g_R\Id+\alpha_RD,
\end{equation}
with real dimensionless $g_{L,R}$ and real $\alpha_{L,R}$ of dimension inverse mass squared. Write $R_0$ for the recoil operator at $\alpha_L=\alpha_R=0$ and $\delta R=R-R_0$. Expanding Eq.~\eqref{eq:positive} to first order in $\alpha_{L,R}$, keeping the order of the factors $D$ and $P_e$, gives
\begin{equation}\label{eq:Rlinear}
\delta R=\mathcal C_S D
+\left(\alpha_L-\frac h2\alpha_R\right)\{P_e,D\}+O(\alpha^2).
\end{equation}
Since $\Tr[\rho\{P_e,D\}]=2\Ree Q_2$, the corresponding change in the dimensionless recoil factor at fixed incident $\rho$ is
\begin{equation}\label{eq:interference}
\begin{aligned}
\delta\Tr(\rho R)={}&2\left(\alpha_L-\frac h2\alpha_R\right)\Ree Q_2\\
&{}+\mathcal C_S S_2+O(\alpha^2),\\
\mathcal C_S={}&\alpha_L(2g_L-hg_R)+\alpha_R(2fg_R-hg_L),
\end{aligned}
\end{equation}
where $S_2=\Tr(\rho D)$. If the same interaction acts on electrons along the path, it adds $\delta H=V(\alpha_L+\alpha_R)D$ to the forward Hamiltonian up to a scalar~\cite{Coloma2023ES}. The incident state entering the recoil rate must then be computed with this potential.

The proportional choice $\alpha_L=\lambda g_L$, $\alpha_R=\lambda g_R$ with known $\lambda\ne0$ gives a way to test the condition $\Ree Q_2=0$ in Eq.~\eqref{eq:ordering_ceiling}. Set $a=g_L^2+fg_R^2-hg_Lg_R$ and $b=2g_L-hg_R$. For the same incident state, the exact tree-level excess over the ordinary recoil rate, in units of $2G_F^2m_e/\pi$, is
\begin{equation}\label{eq:two_bins}
\begin{aligned}
 \Tr[\rho(R-R_0)]&=\lambda b\,\Ree Q_2+a(2\lambda S_2+\lambda^2 S_4),\\
 S_4&=\Tr(\rho D^2).
\end{aligned}
\end{equation}
At fixed incident energy, two recoil-energy bins with linearly independent bin averages of $(b,a)$ determine $\Ree Q_2$ and the combination $2\lambda S_2+\lambda^2S_4$. Extracting these quantities requires calibrated couplings, flux, exposure, and efficiencies, as well as background subtraction and an independent value of $P_{ee}$ for the ordinary rate. The total recoil rate generally remains finite when $\Ree Q_2=0$. Precision predictions also require Standard Model radiative corrections~\cite{Bahcall1995}.

The need for external coupling information can be seen directly in the general model of Eq.~\eqref{eq:G}. Hold $\alpha_L,\alpha_R$ fixed and allow $g_L,g_R$ to vary independently of them. For any shift $c$ that keeps all squared masses nonnegative, the transformation
\begin{equation}\label{eq:couplingdegeneracy}
\begin{aligned}
D&\longrightarrow D+c\Id,\\
g_L&\longrightarrow g_L-\alpha_Lc,\\
g_R&\longrightarrow g_R-\alpha_Rc
\end{aligned}
\end{equation}
leaves both chiral coupling matrices unchanged and therefore preserves the full recoil operator. It also preserves the propagation density matrix by Eq.~\eqref{eq:shift}. If the forward correction is included, its change is proportional to the identity and again drops out of density-matrix evolution. Thus flavor probabilities and recoil spectra admit the same common mass offset when the universal couplings are free. The calibrated proportional model of Eq.~\eqref{eq:two_bins} fixes the couplings along this direction and permits the two-bin inversion.

The two-bin construction uses the mass dependence of the detection vertex with unitary flavor mixing. Nonunitary mixing provides a different modification of neutrino phenomenology~\cite{Antusch2006,Escrihuela2015,Escrihuela2015Erratum}. In the present recoil channel the interference term in Eq.~\eqref{eq:interference} contains only the real quadrature. Reconstructing the winding of the complex response would require independent access to its imaginary part.

\section{Discussion}\label{sec:discussion}
The triangle window and the survival-probability bound involve the same masses and electron-row mixing weights, but constrain different aspects of a zero. The ordering dependence of the bound reflects the relative mass-squared dispersion $\sigma/\mu$, whereas the realization of a zero is a question of propagation dynamics. The window specifies when independent phases can cancel the response at the mass-basis populations of an electron-neutrino source. Matter propagation can change these populations. The bound in Eq.~\eqref{eq:ordering_ceiling} still applies at detection, since it allows all incident density matrices. The continental trajectory nearly saturates that bound even as its root coordinates shift appreciably across the sampled oscillation parameters and density profiles.

The local index and the slope of the survival probability describe different features of the propagation. A nonzero Jacobian allows a simple zero to move under a small, smooth change of parameters. It places no requirement on the slope of $P_{ee}$ measured across different Earth chords. At the continental zero the survival probability is stationary along the fixed trajectory, but a change in chord length also changes the lengths traversed in the crustal layers. This variation accounts for the finite slope in Eq.~\eqref{eq:measuredslope}.

A finite upper limit on $\Ree Q_2$ can also test the ordering bound. Equation~\eqref{eq:ordering_disk} gives $\Ree Q_2\ge\mu p-\sigma\sqrt{p(1-p)}$. At the continental benchmark probability $p=0.88969$, the adopted inverted-ordering parameters require $\Ree Q_2\ge 2.00\times10^{-3}\,\mathrm{eV}^2$ for any incident state and any lightest mass. The minimum occurs at $m_3=0$, since raising $m_3^2$ increases $\mu$ and leaves $\sigma$ fixed. An experimental test would compare an independent upper limit on the real part with this lower bound, accounting for uncertainties in $p$, the oscillation parameters, and the recoil measurement.

A common shift of the squared masses changes $\Ree Q_2$ while preserving all standard flavor probabilities. The two recoil bins in Eq.~\eqref{eq:two_bins} illustrate how to supply the missing information. The calibrated proportional couplings give the real response and the mass-moment combination different recoil-energy dependences. In the general interaction, Eq.~\eqref{eq:couplingdegeneracy} allows independently variable flavor-universal couplings to absorb a common mass shift at fixed recoil spectrum.

\section{Conclusions}
For three-flavor neutrinos, a coherent cancellation of the mass-weighted electron-flavor response can coexist with a large survival probability. The triangle criterion identifies the masses and mixing weights for which such phase configurations exist at fixed source populations. In matter, the projected propagation equation identifies the weighted amplitude with the mass contribution to the local derivative. In the minimal Dirac extension, the real response also enters the interference of Higgs exchange with the charged current in scattering on electrons. An explicit example in layered continental crust reaches a zero at $P_{ee}=0.88969$, close to the maximum of $0.89301$ for the normal-ordering benchmark with $m_1=7$ meV.

At the adopted central oscillation parameters, the maximum electron-flavor probability compatible with a vanishing real part in inverted ordering is about $0.0228$, over all incident states and all nonnegative lightest masses. Interpreting this difference experimentally requires a measurement of the real part in addition to the survival probability. Two recoil bins provide such a test in the calibrated interaction model. In the general recoil model, independently variable flavor-universal couplings can absorb a common shift of the squared masses. Reconstructing the full complex response would also require sensitivity to its imaginary part.

\backmatter
\section*{Statements and Declarations}
\bmhead*{Funding}
No specific funding was received for this study.

\bmhead*{Competing interests}
The authors declare that they have no competing interests relevant to this work.

\bmhead*{Data and code availability}
The numerical data and custom Python scripts supporting the figures, propagation roots, parameter sampling, mass-ordering bounds, recoil calculations, and convergence tests are available from the corresponding author upon reasonable request. The oscillation input parameters for both orderings are publicly available in the NuFIT 6.1 IC24 tables with Super-Kamiokande atmospheric data~\cite{NuFIT61}.

\begin{appendices}
\section{Earth-chord propagation and differentiation}\label{app:earth}
\setcounter{equation}{0}\renewcommand{\theequation}{\Alph{section}\arabic{equation}}
\renewcommand{\theHequation}{appendix.\Alph{section}.\arabic{equation}}
We give the radial mass density $\rho_m$ in $\mathrm{g\,cm^{-3}}$ and the depth $h$ in km. The uniform slab has $\rho_m=2.80$. In the PREM-inspired model, $\rho_m=(2.60,2.90,3.38)$ on the successive depth intervals $[0,12)$, $[12,24.4)$, and $[24.4,\infty)$. In the continental model, $\rho_m=(2.60,2.78,2.95,3.38)$ on $[0,5)$, $[5,15)$, $[15,35)$, and $[35,\infty)$. These profiles are illustrative. The depth-gradient profile (\texttt{smooth} in the code) has $\rho_m=2.62+0.022h$ for $0\le h<12$, $\rho_m=2.884+0.010(h-12)$ for $12\le h<32$, and $\rho_m=3.37$ for $h\ge32$. The chord at the central depth-gradient root reaches a maximum depth of $9.12$ km, within the first branch. Some chords in the parameter sample cross the change of slope at $12$ km. All remain above the density jump at $32$ km.

A chord crosses a spherical interface at depth $h_k$ at distances $s=L/2\pm\sqrt{(R_\oplus-h_k)^2-R_\oplus^2+L^2/4}$ from the source when the radicand is positive. These crossings divide the chord into constant-density segments of lengths $\Delta s_j$. In the mass basis the propagator is
\begin{equation}\label{eq:product}
\begin{aligned}
U_L(L,0)&=U_n\cdots U_1,\\
U_j&=\exp\!\left[-i\kappa\left(\frac{D}{2E}+V_jP_e\right)\Delta s_j\right].
\end{aligned}
\end{equation}
For each segment, write $K_j=-i\kappa H_j\Delta s_j$. The derivative of $U_j$ is $(\partial_L K_j)e^{K_j}$ if $[K_j,\partial_L K_j]=0$. This condition holds when $H_j$ is fixed and only the segment length varies. The general derivative is
\begin{equation}\label{eq:frechet}
\partial_L e^{K_j}=\int_0^1 e^{(1-t)K_j}
(\partial_L K_j)e^{tK_j}\,dt.
\end{equation}
Differentiating the ordered product accounts for the change in each segment length and any explicit dependence of the layer potential on $L$. In the shell models used here, only the lengths vary. For a smoothly varying family of continuous profiles, differentiating the Dyson series gives Eq.~\eqref{eq:chordderivative}. Moving sharp interfaces yield distributional terms away from chord tangencies. At fixed energy, the baseline derivative includes the change in the density profile sampled by the chord.

We diagonalize each $3\times3$ Hermitian layer Hamiltonian and apply the propagators in source-to-detector order. For nonzero dephasing we exponentiate the $9\times9$ Liouvillian in each segment. In the depth-gradient model the root lies at $L_0=681.5701$ km with 24 midpoint slices, $681.5372$ km with 384 slices, and $681.5371$ km with 768 slices. Table~\ref{tab:roots} gives the baselines to $0.1$ km. Grid refinement and the solver residual measure numerical accuracy for the specified inputs. In the shell models we recompute the intersections at each trial baseline, including near tangency, where the chord starts to sample another shell.

\end{appendices}

\end{document}